\documentclass{kluwer}    

\usepackage{epsfig}

\begin{document}                                                                                   
\begin{article}
\begin{opening}         
\title{Extragalactic Astronomy with the VLTI: a new window on the Universe}
\author{Alessandro \surname{Marconi}}  
\author{Roberto \surname{Maiolino}}  
\institute{INAF-Osservatorio Astrofisico di Arcetri, Largo E. Fermi 5,
I-50125 Firenze, Italy}
\author{Romain G. \surname{Petrov}}  
\institute{Universit\'e de Nice Sophia Antipolis UMR 6525, Parc Valrose, 06108 Nice, France}
\runningauthor{Marconi et al.}
\runningtitle{Extragalactic Science with the VLTI}
\begin{abstract}
Interferometry in the optical and near infrared has so far played a marginal
role in Extragalactic Astronomy. Active Galactic Nuclei are the brightest and
most compact extragalactic sources, nonetheless only a very limited number
could be studied with speckle interferometry and none with long baseline
interferometry. The VLTI will allow the study of moderately faint extragalactic
objects with very high spatial resolution thus opening a new window on
the universe.  With this paper we focus on three scientific cases to show how
AMBER and MIDI can be used to tackle open issues in extragalactic astronomy.
\end{abstract}
\end{opening}           

\section{Introduction}  

Optical and near infrared interferometry has so far played a marginal role in
extragalactic astronomy and in particular no multiple aperture observations have
ever been performed. The VLTI sensitivity will allow to
explore spatial scales which have not been probed so far, thus opening a new
window on extragalactic astronomy.
Previous observations of extragalactic sources have been limited to
speckle interferometry of a few of the brightest Active Galactic Nuclei like
NGC 1068
\cite{meab,chelli,afan,witt,weinb},
NGC 1386 \cite{maud}
and NGC 4151 \cite{ebst,aye}.
Even with the use of the 8m telescopes, VLTI observations will be limited
to the brightest sources. The objects with the highest flux within 
the diffraction limit PSF of the 8m telescopes are Active Galactic
Nuclei, which, at least initially, will be the only extragalactic
targets observed.

\begin{figure}
\centering
\epsfig{figure=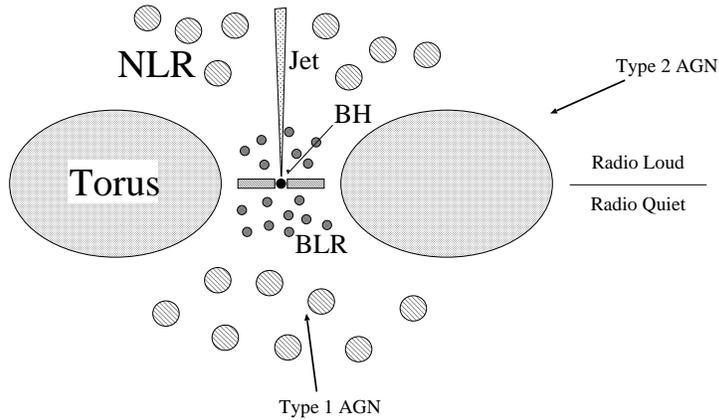,angle=-90,width=0.8\linewidth}
\caption[]{Schematic view of the unified model for AGNs.
The components are not drawn in scale. The BLR (Broad Line Region) size
ranges from a few light days in local Seyfert 1 galaxies to a few light years
in bright quasars. Similarly the obscuring torus size ranges from $\sim 1$ to
$\sim 100$ pc. Finally the NLR (Narrow Line Region) can range from $\sim 100$
pc up to several Kpc.}
\label{fig:agnmod}
\end{figure}

Active Galactic Nuclei are galactic nuclei showing evidence for non-stellar
production of energy. The most stringent constraints come from the emitted
luminosities (L$\sim 10^8-10^{13}$ L$_\odot$), the flat ($L_\nu \sim \nu^{-1}$)
non-stellar spectra extending from radio to $\gamma$-rays, the high
efficiencies of matter-energy conversion ($\simeq 0.1$), the rapid time
variabilities
(observed on scales as short as a few hours in optical and X-rays),
the compact source sizes 
(directly measured in radio sources and smaller than a few light days) and the presence of relativistic jets (e.g.\
\opencite{saasfee}).  Observationally all AGNs can be roughly divided into two
main classes, type 1 and type 2 (e.g.\ Seyfert 1 and 2 nuclei).  The type 1
AGNs show broad ($FWHM>1000$ km/s) permitted emission lines in their optical
and near-IR spectra while type 2 AGNs only show narrow lines ($FWHM<1000$ km/s).
The most widely accepted model (see Fig. \ref{fig:agnmod}) comprises a central
black hole with mass in the range $10^6-10^{10}$ M$_\odot$ surrounded by an
accretion disk that converts gravitational energy into radiation and outflows.
The radiation emitted in the optical - UV - soft X-rays accounts for most of
the AGN bolometric luminosity.  Broad emission lines originate in
small high density gas clouds ($N_\mathrm{e}\sim 10^9$ cm$^{-3}$) 
orbiting around the nuclear source. Variability time scales of
the broad line fluxes and other theoretical arguments suggest that the BLR size
ranges from a few light days for nearby Seyfert 1 galaxies up to a few light
years in the most luminous quasars.  Plasma jets are emitted perpendicular to
the disk. At large radii ($\sim 1-100$ pc), an obscuring torus of cold gas and
dust surrounds the nucleus.
The orientation of this torus relative to the
line of sight naturally accounts for the differences between type 1 and type 2
AGNs (figure \ref{fig:agnmod}; see \opencite{anto} for a review).
Up to now optical and near-IR observations have been able to probe 
spatial scales larger
than the size of the obscuring torus.  Interferometric radio observations are
able to probe smaller scales, but very little
information is available on many constituent of the AGN model,
like the torus or the BLR. Thus, in the following, we will show how the VLTI,
and in particular AMBER,
can be used to tackle open issues on the obscuring torus,
the BLR and the central supermassive Black Hole.

\section{The Obscuring Torus}

The nuclear near-IR luminosity of several AGNs is dominated by emission of hot
dust, close to the sublimation limit, which traces the inner walls of the
circumnuclear molecular torus.  The inner radius of the torus is set by the
minimum distance from the nuclear UV source at which dust can survive against
sublimation, i.e. $R_{in} \simeq 0.2-4\,L_{46}^{1/2}$pc where $L_{46}$ is the
optical-UV luminosity of the AGN in units of $10^{46}$ erg/s \cite{laor}.  The
smallest radius is for large grains and the largest for small grains.  Within
the torus, the absorbing dust reprocesses the optical-UV primary radiation into
the infrared, via heating of the dust grains.  Several authors have modeled the
infrared spectral energy distribution of AGNs and have shown that, indeed, the
nuclear infrared emission can be explained with reprocessing of the nuclear
radiation by dust (e.g.\ \opencite{pier}; \opencite{grana}; \opencite{efst}).
The obscuring dusty medium is
expected to extend from the sublimation radius (less than a pc) up to about 100
pc \cite{grana97,maio95}.

Until a few years ago, the only constraints on the models were the infrared
spectral energy distributions and the upper limits on the size of the nuclear H
and K band sources obtained from ground-based and HST/NICMOS observations (e.g.\
\opencite{thatte}; \opencite{maio98}).
Without strong constraints on the sizes and geometries
of the tori, there is a degeneracy between the model parameters (like dust
composition, distribution etc.) meaning that the same spectra and size
constraints can be adequately explained with very different torus models.  The
radiative transfer models have been recently compared with the IR SED of nearby
Seyfert galaxies and with high angular resolution near- and mid-IR observations
to try to constrain the physics and the geometry of the dusty torus (e.g.\
\opencite{alloin}; \opencite{bock}; \opencite{maio98};
\opencite{rouan}) with the finding that the near- and mid-IR
emitting dust is somewhat more extended than expected by the early models. The
geometry and the physics of the dusty gas in the circumnuclear region of AGN
are thus more complex than assumed so far.
An additional complication is that the properties of dust in AGNs might be
different with respect to the diffuse interstellar medium. More specifically,
\citeauthor{maio01a} \shortcite{maio01a,maio01b} 
have shown that probably the dust in the
circumnuclear region of AGNs is biased in favor of large grains, while
in all models for the molecular torus a ``standard'' Galactic dust mixture has
been assumed. A distribution of dust biased in favor of large grains would
change significantly the distribution of temperatures and also the location of
the dust sublimation radius.
\begin{figure}
\centering
\epsfig{figure=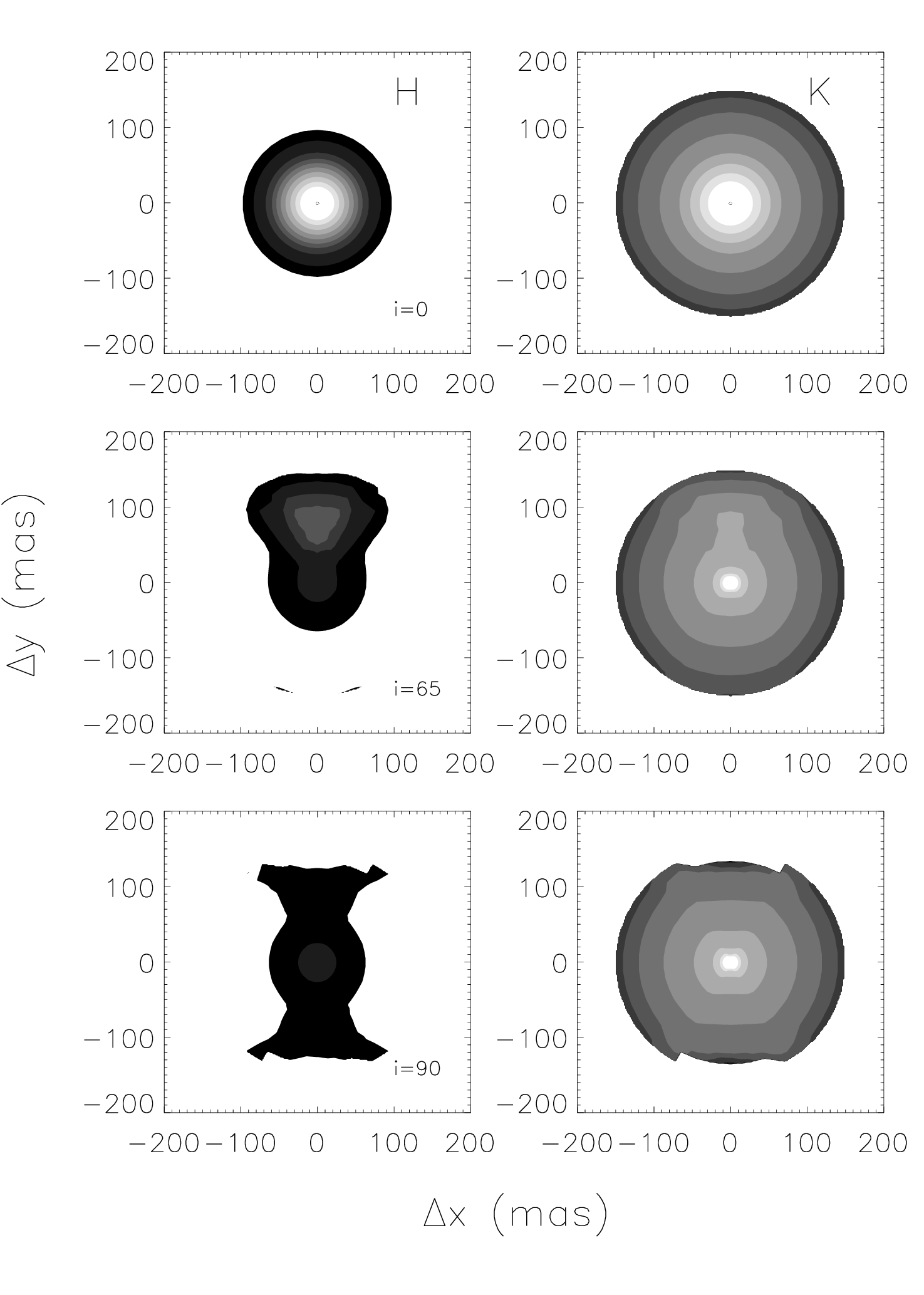,angle=0,width=0.45\linewidth}
\epsfig{figure=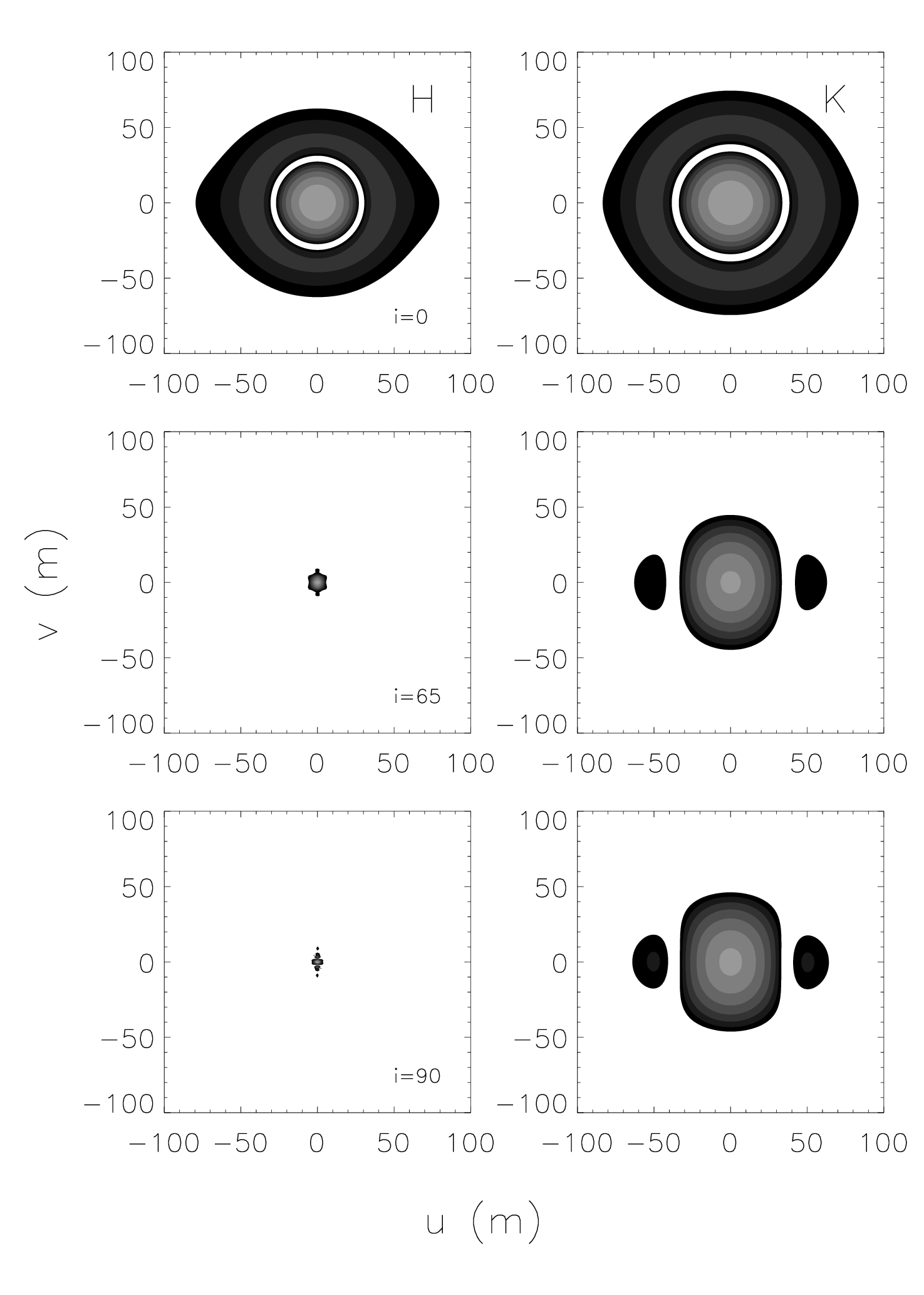,angle=0,width=0.45\linewidth}
\caption[]{Left: Expected surface brightness distributions of H and K emission
for the NGC 1068 torus model by \inlinecite{grana97} seen at different
inclinations with the line of sight. There are eight greyscale levels starting
from -15 with steps of 0.5 in logarithm of erg s$^{-1}$ cm$^{-2}$ hz$^{-1}$
sterad$^{-1}$. Smaller intensity values are darker.
(Model data courtesy of G.L. Granato).
Right: Expected visibilities, $\mid V\mid$.
Notations are as in the previous panel.
There are eight greyscale levels with logarithmic steps of 0.2dex starting from
0.04.}
\label{fig:granato}
\end{figure}

We can consider, for example, the case of the famous Seyfert 2 galaxy NGC 1068
whose luminosity, for an assumed distance of D=14.4 Mpc, is $L=3.7\,10^{45}$
erg/s. The inner radius for a Galactic dust mixture is $0.5L_{46}^{0.5}$
pc, thus the expected inner radius for the NGC 1068 torus is 0.3pc
corresponding to $\simeq 4$ mas. Then, in the model by \inlinecite{grana97},
the torus should have a K-band size of $\simeq 14$ mas FWHM
(18 mas at 10$\mu$m).  These sizes can be directly measured using AMBER and
MIDI at the VLTI. As a zero order approximation, the observed visibilities 
at different points in the uv plane, in different spectral bands
(H, K and 10$\mu$m) can be fit with a ring model to estimate the inner and
outer radii. The measure of the inner radius will directly tell if
the dust composition is as in our galaxy or biased toward larger or smaller
grains.
However one can perform more detailed studies and try to distinguish between
different models. A library of models with different parameters (inner radius,
thickness, inclination, density, optical depth, mixture of dust grains) can be
computed. Then for each model one can compute the expected visibilities and
phases for the points in the uv plane covered by the observations.
By using AMBER with 3 telescopes and the low spectral resolution mode 
(${\cal R}=35$),
a single observation will provide 3 visibilities and one closure phase in any
of the $\sim 30$ spectral channels present in J, H and K.  It is important to
remember that the large wavelength coverage (1.0 to 2.4 $\mu$m) substantially
increases the instantaneous coverage of the uv plane 
which can be further increased taking advantage of Earth
rotation.  MIDI will similarly provide one visibility point per single
observation. Thus, the best model should be able to reproduce at the same time
the AMBER and MIDI observations together with the ir spectra.
Figure \ref{fig:granato} shows the expected spatial distributions
and visibilities of H and K
emission for the NGC 1068 torus model by \inlinecite{grana97}
seen at different inclinations with respect to the line of sight.
For a given geometry, different inclinations produce different
morphologies which can be distinguished with VLTI.

In conclusion, the AMBER and MIDI data will remove the degeneracy among the
model fitting parameters, at variance with previous studies which could compare
models only with the IR spectral energy distribution.

\section{The Broad Line Region of Active Galactic Nuclei}  

The BLR (Broad Line Region) is the region where
the broad (FWHM$>1000$ km/s) permitted lines observed in the spectra of type
1 AGNs originate (e.g.\ \opencite{saasfee}).
Given the small distance from the central super massive black hole (SMBH) the
width of the broad lines is
likely to originate from the gravitational motion of gas clouds around
the SMBH.
So far, the size of the BLR could not be directly measured
and the only available information is provided by the so-called
reverberation mapping technique (e.g.\ \opencite{peter}). The BLR size is
estimated as $c\Delta\tau$ where $\Delta\tau$ is the time lag between the
continuum and line variation.
Clearly this represents an average size
weighted over the BLR geometry and physical conditions. In principle the BLR
geometry and kinematics can be derived from the detailed behavior of the
light curves, but the inversion is not unique mainly because of the
correspondence between the 1-dimensional nature of light curves and the
3-dimensional nature of the BLR.  Also the non optimal time sampling of the
observations strongly reduces the constraints which can be derived.
Suggested spatial distribution of the BLR clouds are spherical, disk-like or
conical. Dynamically, the BLR might be dominated by gravitational motions
(either a virialized system with chaotic motions or a disk in keplerian
rotation), or might be part of a radiation pressure driven outflow, or of an
inflow. Obtaining a direct measure of the BLR size and
constraining its morphology and
kinematics is fundamental in order to understand its origins and relationship
with AGN activity, and to measure the mass of the SMBH and verify reverberation
mapping techniques.  BLR sizes determined with the reverberation mapping have
been found to correlate with the quasar luminosity. A recent estimate by
\inlinecite{kaspi} gives:
\begin{equation}
R_{BLR} \sim 33 \left(\frac{\lambda L_\lambda (5100\mathrm{\AA})}
{10^{44}\,erg\,s^{-1}}\right)^{0.7}
\mathrm{light~days}
\end{equation}
where $L_\lambda$ is the rest-frame monochromatic luminosity at 5100 \AA.
Assuming a standard cosmology ($H_0=70$ km/s/Mpc, $\Omega_M=0.3$ and
$\Omega_\Lambda=0.7$), $R_{BLR}$ can be translated into an apparent size in the
plane of the sky obtaining for instance $R_{BLR}=0.03$ mas for a $L=10^{12}$
L$_\odot$ quasar at $z=0.1$ (we have assumed $\lambda L_\lambda
(5100\mathrm{\AA})$ = L/15).  For $L=10^{13}$ L$_\odot$ and $z=0.5$,
$R_{BLR}=0.04$ mas.  These numbers imply that BLRs are not resolvable with the
VLTI. However, when a source is not resolved by an interferometer, the
differential phase measures the displacement of the photocenter with wavelength
along the baseline direction.
This is valid even when this displacement 
is much smaller than the standard resolution limit of the interferometer.
The position of the photocenter with wavelength
is a powerful constraint for the geometry and kinematics of the BLR and can be
obtained by taking advantage of the AMBER spectral resolution.
\begin{figure}  
\centering
\epsfig{file=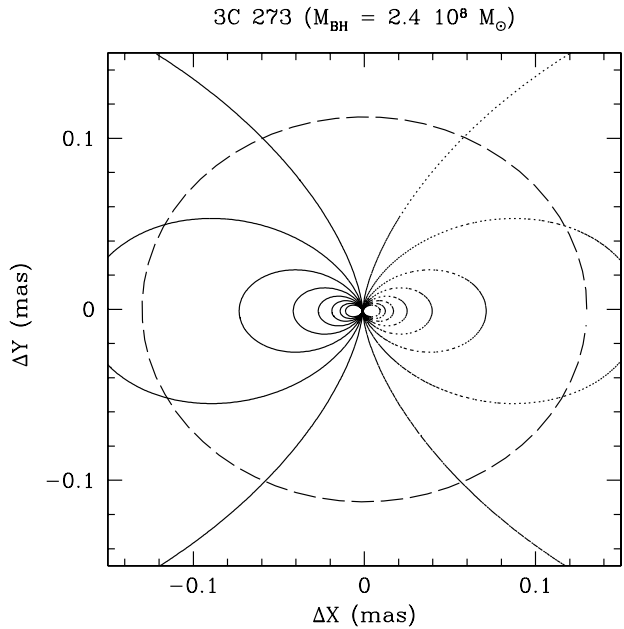,width=0.7\linewidth}
\caption[]{Isovelocity contours for the BLR of 3C273. `Red' velocities are represented with a solid line, `blue' velocities with a dotted line. Contours go
from 400 to 3200 km/s with steps of 400 km/s. The dashed line is the BLR radius measured with reverberation mapping. The assumed disk inclination with respect to the line of sight is 30 deg.}
\label{fig:blr}
\end{figure}

For example, one can consider the case of the famous quasar 3C273.  Its BLR
radius measured with reverberation mapping by \inlinecite{kaspi},
$R_{BLR}=387\pm55$ lt-days corresponds to 0.13 mas (at the distance of 3C 273,
1 mas corresponds to 2.6 pc) and the inferred BH mass is $\sim 2.4 \times 10^8$
M$_\odot$.  Figure \ref{fig:blr} shows the expected isovelocity contours with
steps of 400 km/s assuming that the BLR is in a Keplerian disk and that it is
inclined of 30 deg with respect to the line of sight.  Using AMBER with Medium
spectral resolution (${\cal R}=750$), one can measure the differential phase of
the broad Pa$\alpha$ in the K band and obtain information in 400 km/s velocity
bins.  At 2.2$\mu$m, with a projected baseline of 80 m,
a photocenter displacement of 0.13 mas will result in a phase shift
on the line of $\phi\sim 0.15$ rad. Since in each spectral channel we expect
$\sim 300$ Pa$\alpha$ photons/sec, 
the expected accuracy which can be reached in 1 h of observation is
$\sigma_\phi\sim 0.002$ rad
($\sim 2\, \mu$arcsec in K with a 80 m baseline),
thus allowing to measure the photocenter displacement.  Since a
given measure with two telescopes yields mainly one component of the
photocenter displacement, in order to obtain a full map of the photocenter
displacement one should measure the differential phase using observations from
two perpendicular baselines.

The differential
phase in each velocity bin will provide the position of the photocenter of
the part of the disk which is delimited by two contiguous contours in 
figure \ref{fig:blr}.  In this
case, the photocenters will be aligned along the line of nodes of the disk
and, combining positional and velocity information, one can directly 
measure the mass
of the central BH.  Conversely, if the BLR is composed by a spherically
symmetric and isotropic ensemble of clouds the photocenters will all
coincide with the position of the continuum photocenter and no differential
phase will be measured.  
Other locations of the photocenters will result in more
complicated geometries but, in any case, 
the amplitude of the photocenter displacement  
is directly related to the angular size of the BLR.
If we are able to establish a relationship between the differential
interferometry angular radius and the reverberation mapping linear radius, we
can get a direct measurement of the AGN distance.  Since, in the best case,
this kind of measurement can be made on a magnitude $V\sim 20$ quasar when
there is a nearby reference star, this could eventually lead to a new and
independent technique to measure cosmological distances.

\section{The central Black Hole}

It is now clear that a large fraction of local galaxies, if not all, contain a
MBH (e.g.\ \opencite{kg}).  Moreover, there is a suggestion in the
data that the hole mass is proportional to the mass (or luminosity) of the host
spheroid (e.g.\ \opencite{mf01} and references therein).  Recently
\inlinecite{fm00} and \inlinecite{geb00} have shown that a
tighter correlation holds between the BH mass and the velocity dispersion of
the bulge.  Clearly, any correlation of black hole and spheroid properties
would have important implications for theories of galaxy formation in general,
and bulge formation in particular.
These results have prompted a frantic attempt to understand the physical origin
of the observed correlations between BH mass and galaxy properties (e.g.\
\opencite{hk}; \opencite{cava}) and to put constraints
on galaxy and BH formation models.

However, there is a caveat which must be taken into account: the so called
massive black holes in galactic nuclei are in reality massive dark objects
because, at the moment, there is no conclusive proof that they are indeed black
holes.  The massive dark objects are point-like at the resolution of the
observations, which, at most, are of the order of 0.1". Combining this with
typical galaxy distances of 10-20Mpc, one derives that the density of the
massive dark objects are not so high that the only plausible alternative is a
BH: indeed in most cases they are even smaller than the densities observed in
core collapsed globular cluster ($\rho\sim 10^7\,M_\odot\,pc^{-3}$) which are
the most dense star clusters known.
Therefore the possibility that these dark masses could be clusters of "dark"
objects (e.g.\ neutron stars, planets etc.) cannot be excluded. Two notable
exceptions are our galactic center 
($\rho\sim 10^{12.6}\,M_\odot\,pc^{-3}$, \opencite{genzel}) and NGC 4258
($\rho\sim 10^{9.5}\,M_\odot\,pc^{-3}$,
\opencite{4258}) where, given the high densities of the dark matter distribution, the only possibility is that of a massive black
hole.  In the first case the key point is the closeness (just 8 kpc) while in
the second case the high spatial resolution is obtained with VLBI observations
of water masers.  Unfortunately our galactic center is unique and galaxies with
water masers observable with VLBI are just a handful and none of
them provides as good an evidence as NGC 4258 \cite{masers}.
\begin{figure}  
\centering
\epsfig{file=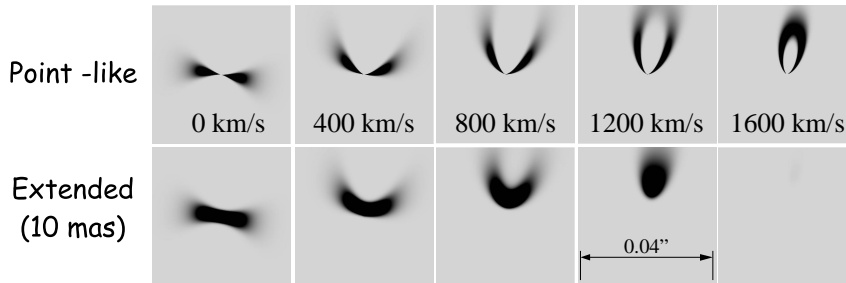,angle=-90,width=0.95\linewidth}
\caption[]{Intensity distribution from a rotating disk
in various velocity ranges (toplabels; km/s) expected 
from a point like mass (top) and a mass extended over 10mas (bottom). 
The case considered is that of Centaurus A where a dark mass
of $10^8\, M_\odot$ is detected from the rotation of a gaseous disk inclined
by $35^\circ$ with the line of sight.
The images are 0.04" x 0.04". North is up and East is left.}
\label{fig:cena}
\end{figure}

Using the spectral resolution of AMBER it is possible to constrain the size
of the dark mass distributions which are found at the galaxy centers to verify
if they are indeed black holes.  We can consider for example the case of
Centaurus A \cite{marconi}.  In Centaurus A the dark mass is $\sim 10^8
M_\odot$ and if it can be constrained
within 10mas (i.e. 0.17pc) the density will
be $2\times 10^{10} M_\odot pc^{-3}$. Centaurus A will thus
provide a better case
for a BH than NGC 4258.
Centaurus A has an ionized gas disk rotating around the putative BH. This disk
is a strong emitter in H and Fe lines and can be observed with AMBER in
Pa$\beta$ and [FeII] 1.26$\mu$m with medium spectral resolution (${\cal
R}=750$, 400 km/s spectral channels).  On the spatial scales accessible to
AMBER, this disk will rotate with velocities of $\sim 0-3000$km/s and the
regions at different velocities will have different morphologies if the mass is
point like or extended (see Fig.  \ref{fig:cena}). The extreme case is for the
velocity bin at 1600 km/s where no disk emission is expected if the mass is
extended over 10 mas.  The visibility points and differential phases obtained
at several locations in the uv plane for each spectral channel can be compared
with models of disks rotating around a point mass (BH) or around an extended
mass distribution (several different distributions can be tested).  By fitting
the model to the data one can then determine the free parameters (total mass,
PA of disk line of nodes, disk inclination) as well as constrain the region
were the dynamical mass must be distributed.

\section{Conclusions}

The VLTI will allow the study of moderately faint extragalactic objects with
very high spatial spatial resolution thus opening a new window on the universe.
To show the potentiality of high spatial resolution interferometric
observations we have presented three scientific cases which will be tackled
with AMBER and MIDI.  The combined use of AMBER and MIDI will make possible to
spatially resolve the obscuring torus of Active Galactic Nuclei and to solve
the degeneracies in torus models.  The unique spectroscopic capabilities of
AMBER will be able to spatially resolve the BLR and to constrain its geometry
and kinematics.
By relating the BLR sizes measured with reverberation mapping and
those measured with differential interferometry one might directly measure
cosmological distances.
Finally, it will also be possible to constrain the size of the dark
matter distributions found in galaxy nuclei to test if they are really
supermassive black holes.
%

\end{article}

\begin{thebibliography}{}
\bibitem[\protect\citeauthoryear{Afanas'ev et al.}{1992}]{afan}
Afanas'ev, V.~L., I.~I.~Balega, V.~G.~Orlov, \& V.~A.~Vasiuk 1992,
A\&A 266, 15-20 
\bibitem[\protect\citeauthoryear{Alloin et al.}{2000}]{alloin}
Alloin, D., E.~Pantin, P.~O.~Lagage, \& G.~L.~Granato 2000,
A\&A 363, 926-932
\bibitem[\protect\citeauthoryear{Antonucci}{1993}]{anto} Antonucci, R. 1993,
ARA\&A 31, 473-521
\bibitem[\protect\citeauthoryear{Ayers et al.}{1990}]{aye} Ayers, G.~R., J.~Benson, 
K.~Carels, et al.\  1990,
ApJ 360, 471-473
\bibitem[\protect\citeauthoryear{Blandford, Netzer \& Woltjer}{1990}]{saasfee}
Blandford, R.~D., H.~Netzer, L.~Woltjer 1990,
Saas-Fee Advanced Course 20,~Lecture Notes 1990, XII, 280 pp.~97 
Springer-Verlag
\bibitem[\protect\citeauthoryear{Bock et al.}{2000}]{bock} Bock, J.~J., 
G.~Neugebauer, K.~Matthews, et al.\  2000,
AJ 120, 2904-2919
\bibitem[\protect\citeauthoryear{Braatz et al.}{1997}]{masers} Braatz, J.~A., 
A.~S.~Wilson, \& C.~Henkel 1997,
ApJS 110, 321
\bibitem[\protect\citeauthoryear{Cavaliere \& Vittorini}{2002}]{cava} Cavaliere, 
A.~\& V.~Vittorini 2002,
ApJ 570, 114-118
\bibitem[\protect\citeauthoryear{Chelli et al.}{1987}]{chelli} Chelli, A., 
I.~Cruz-Gonzalez, L.~Carrasco, \& C.~Perrier 1987,
A\&A 177, 51-62
\bibitem[\protect\citeauthoryear{Ebstein et al.}{1989}]{ebst} Ebstein, S.~M., 
N.~P.~Carleton, \& C.~Papaliolios 1989,
ApJ 336, 103-111
\bibitem[\protect\citeauthoryear{Efstathiou \& Rowan-Robinson}{1995}]{efst} 
Efstathiou, A.~\& M.~Rowan-Robinson 1995,
MNRAS 273, 649-661
\bibitem[\protect\citeauthoryear{Ferrarese \& Merritt}{2000}]{fm00} Ferrarese, 
L.~\& D.~Merritt 2000,
ApJ 539, L9-L12
\bibitem[\protect\citeauthoryear{Gebhardt et al.}{2000}]{geb00}
Gebhardt, K.~et al.\ 2000,
ApJ 539, L13-L16
\bibitem[\protect\citeauthoryear{Genzel et al.}{2000}]{genzel}
Genzel, R., C.~Pichon, A.~Eckart, et al.\  2000,
MNRAS 317, 348-374
\bibitem[\protect\citeauthoryear{Granato \& Danese}{1994}]{grana} Granato, G.~L.~\& 
L.~Danese 1994,
MNRAS 268, 235
\bibitem[\protect\citeauthoryear{Granato et al.}{1997}]{grana97} Granato, G.~L., 
L.~Danese, \& A.~Franceschini 1997,
ApJ 486, 147
\bibitem[\protect\citeauthoryear{Haehnelt \& Kauffmann}{2000}]{hk} Haehnelt, 
M.~G.~\& G.~Kauffmann 2000,
MNRAS 
318, L35-L38
\bibitem[\protect\citeauthoryear{Kaspi et al.}{2000}]{kaspi} Kaspi, S., P.~S.~Smith, 
H.~Netzer, et al.\  2000,
ApJ 533, 631-649
\bibitem[\protect\citeauthoryear{Kormendy \& Gebhardt}{2001}]{kg} Kormendy, J.~\& 
K.~Gebhardt 2001,
Proc. of 20th Texas Symposium, 363
\bibitem[\protect\citeauthoryear{Laor \& Draine}{1993}]{laor} Laor, A.~\& 
B.~T.~Draine 1993,
ApJ 402, 441-468
\bibitem[\protect\citeauthoryear{Maiolino \& Rieke}{1995}]{maio95} Maiolino, R.~\& 
G.~H.~Rieke 1995,
ApJ 454, 95
\bibitem[\protect\citeauthoryear{Maiolino et al.}{1998}]{maio98} Maiolino, R., 
A.~Krabbe, N.~Thatte, \& R.~Genzel 1998,
ApJ 493, 650
\bibitem[\protect\citeauthoryear{Maiolino et al.}{2001a}]{maio01a} Maiolino, R., 
A.~Marconi, M.~Salvati, et al.\  2001a,
A\&A 365, 28-36
\bibitem[\protect\citeauthoryear{Maiolino et al.}{2001b}]{maio01b} Maiolino, R., 
A.~Marconi, \& E.~Oliva 2001b,
A\&A 365, 37-48
\bibitem[\protect\citeauthoryear{Marconi et al.}{2001}]{marconi} Marconi, A., 
A.~Capetti, D.~J.~Axon, et al.\ 2001,
ApJ 549, 915-937
\bibitem[\protect\citeauthoryear{Mauder et al.}{1992}]{maud} Mauder, W., 
I.~Appenzeller, K.-H.~Hofmann, et al.\   1992,
A\&A 264, L9-L12
\bibitem[\protect\citeauthoryear{Meaburn et al.}{1982}]{meab} Meaburn, J., 
B.~L.~Morgan, H.~Vine, et al.\  1982,
Nature 296, 331
\bibitem[\protect\citeauthoryear{Merritt \& Ferrarese}{2001}]{mf01} Merritt, D.~\& 
L.~Ferrarese 2001,
ASP Conf.~Ser.~249, 335
\bibitem[\protect\citeauthoryear{Miyoshi et al.}{1995}]{4258} Miyoshi, M., J.~Moran, 
J.~Herrnstein, et al.\  1995,
Nature 373, 127
\bibitem[\protect\citeauthoryear{Peterson}{1994}]{peter} Peterson, B.~M.\ 1994
ASP Conf.~Ser.~ 69, 1
\bibitem[\protect\citeauthoryear{Pier \& Krolik}{1993}]{pier} Pier, E.~A.~\& 
J.~H.~Krolik 1993,
ApJ 418, 673
\bibitem[\protect\citeauthoryear{Rouan et al.}{1998}]{rouan} Rouan, D., F.~Rigaut, 
D.~Alloin,  et al.\   1998,
A\&A 339, 687-692
\bibitem[\protect\citeauthoryear{Thatte et al.}{1997}]{thatte} Thatte, N., 
A.~Quirrenbach, R.~Genzel, et al.\  1997,
ApJ 490, 238
\bibitem[\protect\citeauthoryear{Weinberger et al.}{1999}]{weinb} Weinberger, A.~J., 
G.~Neugebauer, \& K.~Matthews 1999,
AJ 117, 2748-2756
\bibitem[\protect\citeauthoryear{Wittkowski et al.}{1998}]{witt} Wittkowski, M., 
Y.~Balega, T.~Beckert, et al.\  1998,
A\&A 329, L45-L48
\end{thebibliography}
\end{document}